\documentclass[pra,showpacs,twocolumn]{revtex4-1}
\usepackage{bm}
\usepackage{amsmath}
\usepackage{amssymb}
\usepackage{epsfig}

\def\be{\begin{equation}}
\def\en#1{\label{#1}\end{equation}}
\def\d{\dagger}

\newcommand{\eqb}{\begin{eqnarray}}
\newcommand{\eqe}{\end{eqnarray}}

\begin{document}
\title{Generators of  nonclassical states by combination of the linear coupling of boson modes,
Kerr nonlinearity and the strong linear losses }

\author{ V. S. Shchesnovich$^1$ and D.  Mogilevtsev$^2$}
\affiliation{${}^1$Centro de Ci\^encias Naturais e Humanas, Universidade Federal do
ABC, Santo Andr\'e,  SP, 09210-170 Brazil\\
$^2$Institute of Physics, Belarus National Academy of Sciences, F.Skarina Ave. 68,
Minsk 220072 Belarus}

\begin{abstract}

We show that  the generators of   quantum states of light can be built by employing
the Kerr nonlinearity, a strong linear absorption or losses and the linear coupling
of optical modes. Our setup can be realized, for instance,  with the use of the
optical fiber technology. We consider in detail the simplest cases of three and
four coupled modes, where a strongly lossy   mode is linearly coupled to other
linear and nonlinear modes. In the three-mode design, our scheme  emulates the
third-order nonlinear absorption, allowing  for generation of the single photon
states,  or   the two-photon absorption allowing to generate  the phase states. In
the four-mode design, the scheme emulates a non-local absorption which produces an
entangled state  of  two uncoupled  modes. We also note that in the latter case and
in the case of the phase states generation the output state is in the linear modes,
which prevents its subsequent degradation by  strong losses accompanying a strong
Kerr nonlinearity.

\end{abstract}
 \maketitle

\section{Introduction}
Robust, efficient and reliable generators of  nonclassical states of light are
needed for the current and possible future technology. Nonclassical states are the
working tools  of quantum information and computation \cite{IC}, besides other
applications in  quantum diagnostics and tomography, metrology, communications,
etc. An ideal quantum state generator can be thought as  a device producing a given
quantum state by pushing a button, i.e. deterministically. It is known that a
nonlinear media can be used to produce a nonclassical state of light from a
classical (i.e. coherent) one \cite{TF}, for instance, the $n$-photon Fock state
can be obtained, at least in principle, in this way \cite{KCH}. However, the
experimental realization of the deterministic nonclassical state generators faces
such great difficulties that the  ``determinism" requirement is currently dropped
(it is pertinent to note that the overall efficiency of one of the most promising
candidates for the ``on-demand" single photons  generation, the color centers in
diamond, does not exceed a few percent \cite{CSD}). Instead, one has  to be
currently  satisfied with the conditionally generated nonclassical states. For
instance,   the generation of the single photon states \cite{TF,LO} is currently
based on producing the twin photon pairs by the parametric down-conversion  or the
heralded single photons generated in the mesoscopic $pn$ junctions \cite{pnT,pnE}.

The main difficulty of the deterministic nonclassical states generation using a
nonlinear media is the losses, since a strong nonlinearity required for such a
generator  is inevitably accompanied by a strong absorption  (in particular, this
is the case of the nonlinear glasses used for fabrication of the nonlinear optical
fibers). On the other hand,  tailoring of a classical   state to a nonclassical one
by using the nonlinear absorption was also considered. Thus  the dissipation,
considered usually as a destructive effect  to the fragile quantum features (e. g.
the so-called ``Schr$\mathrm{\ddot o}$dinger cat" quantum state is a proverbial
example of this \cite{SCAT}), can be used also to  drive the system into a
nonclassical state (this is the essence  of the ``quantum state protection''
\cite{DAV}). A wide variety of initial mixed states (or even an arbitrary mixed
state) can be transformed into a desired pure state in this way, thus providing an
extreme robustness of the scheme \cite{KRAUSS}. The artificially designed
dissipative systems were also shown to be the means of the quantum computation
\cite{VWC}.

The problem is that the practical realization of a given nonlinear
loss channel for photons is rather difficult. One can engineer the
dissipation needed for generating of the nonclassical states more
or less freely  working with the  ions in a magnetic trap
\cite{ZOLL,WINL}, or with the atoms in the optical lattices
\cite{ZWER}. However, designing similar reservoirs for photons is
a much more challenging task, where even the theoretical schemes
are mostly lacking. A notable exception is a couple of schemes
proposed recently for generating of the phase states $(|0\rangle
+e^{i\phi}|1\rangle)/\sqrt{2}$ \cite{TFAb,TFComm,TFRepl} and the
single-photon states \cite{NlAbs}. Alas, they still require the
very specific nonlinear dissipation (for example, the third-order
nonlinear absorption is needed for the scheme of Ref.
\cite{NlAbs}). Remarkably, in a recent work \cite{MS} we have
shown that  one can emulate the needed nonlinear absorption by a
simple design based on the optical coupler.

The purpose of the present work is to study in a more general setting  the
potential of our method of the quantum state generation,  employed in Ref.
\cite{MS}. The central idea behind our scheme is the emulation of the needed
nonlinear absorption by using a combination of the linear absorption, the Kerr-type
nonlinearity and the mode coupling. It can be explained as follows. Consider  a
strongly lossy local mode linearly coupled to several other local modes. On the
short evolution time (or propagation length), inversely proportional to the strong
absorption, the lossy mode is emptied, whereas the rest of the modes are still
significantly  populated (curiously, this  is due to the quantum Zeno effect
\cite{Zeno}). Then the evolution can again be represented as that of a single lossy
mode coupled to other modes but now in a coherent basis (i.e.  a superposition of
the local modes). Since some of the local modes are nonlinear -- the usual Kerr
nonlinearity is assumed --  the  coupling of this new lossy mode to other coherent
modes is now nonlinear. When the coherent lossy mode is also emptied, the dynamics
of the remaining   coherent modes is governed by the nonlinear absorption, which
can be tailored by selecting the particular setup of mode coupling. We show that,
additionally to the single-photon states, also the phase states and the entangled
states can be  generated by our method. Moreover, output states in the second and
third aforementioned examples are  localized in the linear modes of the system,
thus preventing the degradation of the output quantum state by the strong losses,
always accompanying a strong Kerr nonlinearity.

The paper is organized as follows. In section \ref{sec2} we describe   the general
scheme for the quantum state generation. We focus on  the  three coupled modes in
section \ref{sec3}, where  we consider two nonlinear side modes, subsection
\ref{sec3a},  and one nonlinear and one linear side modes,  subsection \ref{sec3b},
coupled to a common lossy mode. Moreover, we study the generation of multi-photon
Fock states by this scheme in subsection \ref{sec3c}. In section \ref{sec4} the
case of four coupled modes is considered. In the Conclusion we outline the
perspectives of the method.

\section{The general scheme for emulation of nonlinear losses}
\label{sec2}

Our  design for  emulation  of  nonlinear absorption  involves a boson mode subject
to strong linear losses and coupled to other linear and nonlinear boson modes.  We
assume that the dynamics can be described by the master equation in the standard
Lindblad form \cite{ME}. For the system with Hamiltonian $H$ and a strongly lossy
mode $b$, with the loss rate $\Gamma$, the evolution of an arbitrary operator $
A(t)$ is given by the equation
\be
\frac{d
A}{dt} = i[H,A] + \Gamma\mathcal{D}[b^\d]A,
\en{EQ1}
where $t$ is the evolution variable which will be called ``time'' for below (we use
dimensionless variables), $\mathcal{D}[b^\d] A = b^\d A b - \frac12\{b^\d b, A\}$
(here and below the capital letters   denote the time-dependent operators, while
the small letters are used for the time-independent boson mode operators). We
assume that the Hamiltonian $H$ can be expanded in the powers of the lossy mode as
follows
\be H = \sum_{\alpha,\beta=0}{H}^{(\alpha,\beta)}(b^\d)^\alpha b^\beta.
\en{EQ2}
Here the coefficients ${H}^{(\alpha,\beta)}$ are operators in all other modes of
the system. We limit our consideration to the Kerr nonlinearity, thus $\alpha,\beta
\le 2$ (note, however, that our method works for  arbitrary nonlinearity).

Any time-dependent observable $A(t)$ can always be expanded in the infinite power
series with respect to $b^\d$ and $b$: \be A(t) =
\sum_{\alpha,\beta\ge0}A^{(\alpha,\beta)}(t)(b^\d)^\alpha b^\beta,
\en{EQ3} where the coefficients $A^{(\alpha,\beta)}(t)$ are
time-dependent functions  of   the creation and annihilation operators of the other
modes of the system.

For strong losses, i.e. $\Gamma \gg ||H^{(\alpha,\beta)}||$, for the evolution
times $t\gg 1/\Gamma$ the mode $b$ is  almost empty and  one can adiabatically
eliminate it from the operator equation (\ref{EQ1}). The key observation is that
the leading order term in the expansion, i.~e. $A^{(0,0)}(t)$, is coupled by the
Hamiltonian of Eq. (\ref{EQ2}) to just a few higher-order terms. The latter can be
easily found in the explicit form in the adiabatic approximation. Inserting Eqs.
(\ref{EQ2}) and (\ref{EQ3}) into Eq. (\ref{EQ1}) we have
\[
\frac{d A^{(0,0)}}{dt} = i[H^{(0,0)},A^{(0,0)}] - iA^{(0,1)}H^{(1,0)} +
iH^{(0,1)}A^{(1,0)}
\]
\be +2iH^{(0,2)}A^{(2,0)} - 2iA^{(0,2)}H^{(2,0)}. \en{EQ4}
Observing that for large times, $t\gg \Gamma$, the term $A^{(\alpha, \beta)}(t)$,
with $\alpha+\beta>0$, decays as $\exp\{-\Gamma(\alpha+\beta)t/2\}$  we   retain in
the evolution equation for this  term  the only non-decaying term and the  term
multiplied by a large parameter  (the decay rate $\Gamma$)
\be \frac{d
A^{(\alpha,\beta)}}{dt} \approx i[H^{(\alpha,\beta)},A^{(0,0)}] -
\frac{\Gamma(\alpha +\beta)}{2} A^{(\alpha, \beta)},
\en{EQ5}
which results in
\be A^{(\alpha,\beta)}(t) \approx
\frac{2}{\Gamma(\alpha+\beta)}[H^{(\alpha,\beta)},A^{(0,0)}(t)],\quad
\alpha+\beta>0.
\en{EQ6}
Finally, inserting the result (\ref{EQ6}) into Eq. (\ref{EQ4}), we arrive at the
master equation with   \textit{generally nonlinear} artificial losses
\be
\frac{d
A^{(0,0)}}{dt} \approx i[H^{(0,0)},A^{(0,0)}] +
\frac{4}{\Gamma}\sum_{\alpha=1,2}\mathcal{D}[H^{(0,\alpha)}]A^{(0,0)},
\en{EQ7}
where the Lindblad generator $\mathcal{D}[\cdot]$ is defined below Eq.~(\ref{EQ1}).
Eq. (\ref{EQ7}) has the decay rate in the usual form of  the quantum Zeno effect,
i.e. the actual losses are inversely proportional to the bare loss rate $\Gamma$
\cite{Zeno}.

Let us make the following observations about Eq. (\ref{EQ7}). Though we have
assumed that the terms $H^{(0,1)}$ and $H^{(0,2)}$ in the expansion (\ref{EQ2}) of
the Hamiltonian involve the nonlinearities, the strongly lossy mode is not required
to be nonlinear (in fact, due to the strong loss, its nonlinearity is always
negligible). Moreover, it is only linearly coupled to other modes of the system.
Since Kerr nonlinearities are always diagonal in the spatially local boson basis,
one needs the coupling to more than one mode, which makes such coupling diagonal in
the \textit{non-local} coherent basis and allows to emulate some nonlinear
absorption. Thus, a linear coupling of a strongly lossy mode to two or more other
modes, at least one of which is nonlinear, leads in the second adiabatic reduction
to emulation of a nonlinear (coherent) lossy mode $b$ coupled to other nonlinear
(coherent) modes.

We should mention the conditions which must be imposed for the adiabatic
elimination of the lossy modes (in the two stages of the scheme) and the more
important  condition of the scheme efficiency related to the existence of the
natural or induced (see below) linear losses. These conditions are particular to
each realization  of the general scheme and are given below for each example. We
only mention that the first type of conditions are not actually necessary but only
sufficient for the emulation of the nonlinear absorptions, but they  allow an
analytical treatment via the adiabatic elimination,  as outlined above. The second
type is more important, since, as we will see below, the linear losses, even not
naturally present in the output modes, appear  due to coupling to the lossy modes
though with a significantly reduced rate (as in the case of the phase state
generation, section \ref{sec3b}, and the generation of the entangled state, section
\ref{sec4}). Moreover, both types of conditions point to the conclusion that a
strongly asymmetric modal coupling is needed for the design of a specific nonlinear
dissipation and the  localization of the output nonclassical state in \textit{the
local modes}.

Finally, the   system described   by Eq. (\ref{EQ7}) can be realized using a fiber
coupler or multi-core nonlinear fiber, where the optical mode of the central core
is linearly coupled to the optical modes of the side cores/fibers, where some of
the latter are made of  the nonlinear media \cite{MS}. In this case, the  cores
have to be the single-mode ones, however,  the consideration can be generalized to
the multi-mode case as well. Another possibility is to implement the
electromagnetically induced transparency (EIT) media, where one can achieve a very
high Kerr nonlinearity \cite{kerr} (termed  even the ``giant Kerr nonlinearity").
Co-directional modes travelling through such a media can interact simultaneously
and resonantly with the strongly damped emitters, in this way realizing the
correlated loss of Eq.~(\ref{EQ2}). A variation of this scheme is the propagation
of several differently polarized modes in a nonlinear fiber with a strongly
dissipative impurities (atoms, quantum dots, etc.) present  in the core. When the
average distance between impurities is much larger then the pulse length, the
correlated modal loss can be achieved.

\section{Nonclassical state generation by the three mode coupler}
\label{sec3}

Let us see how the above ideas  can be applied for the case of three interacting
modes. Recently, we have considered some particular realizations of this system
\cite{MS,SM}. Here we analyze in detail the possibilities of  the three-mode
arrangement for generation of the nonclassical states, discussing the symmetrical
and asymmetrical cases and the possibility of localization of the generated state
in the mode with low   losses. We consider the following Hamiltonian
\be
H_3 = \left[(g_1 a_1
+ g_2 a_2)a^\d_0 + h.c.\right] + \kappa_1 (a_1^\d)^2 a_1^2+\kappa_2 (a_2^\d)^2
a_2^2,
\en{EQ8}
where the strong loss is in  mode $a_0$ with the loss rate $\Gamma_0$ such that
$\Gamma_0 \gg G=\sqrt{g_1^2 +g_2^2}$ and $\Gamma_0 \gg |\kappa_{1,2}|$ (here
$g_{1,2}$ are real). The  master equation for an observable
$A=A(t,a^\dag_0,a_0,a^\dag_1,a_1,a^\dag_2,a_2)$ of the whole system reads
\be
\frac{d A}{dt} = i[H_3,A] + \sum_{j=0}^2\Gamma_j\mathcal{D}[a^\d_j]A.
\en{EQ9}
The first reduction of Eq. (\ref{EQ9}),  given by Eq. (\ref{EQ7}) with the
expansion in Eq. (\ref{EQ3})  in powers of $a^\d_0$ and $a_0$, features $H^{(0,1)}
= \sum\limits_{j=1,2}g_j a^\d_j$ and $H^{(0,2)}=0$, and emulates the lossy coherent
mode $b_1$ (with the linear losses rate $4G^2/\Gamma_0$) coupled to another
coherent mode $b_2$. The new modes are given by the following rotation of
local modes ($q = g_1/g_2$) \be \left(\begin{array}{c} b_1\\
b_2 \end{array}\right) = \left(\begin{array}{cc}
\frac{q}{\sqrt{1+q^2}} & \frac{1}{\sqrt{1+q^2}} \\
-\frac{1}{\sqrt{1+q^2}} & \frac{q}{\sqrt{1+q^2}}\end{array}\right)\left(\begin{array}{c} a_1\\
a_2 \end{array}\right).
\en{EQ10}
Eq. (\ref{EQ7}) now involves  the  Lindblad terms for modes $b_{1,2}$ and the
Hamiltonian describing the interaction of  modes $b_1$ and $b_2$, which are derived
from Eq. (\ref{EQ9}):
\[
\sum_{j=1,2}\Gamma_j\mathcal{D}[a^\d_j]A =
\sum_{j=1,2}\biggl\{\frac{q^2\Gamma_{j}+\Gamma_{j^\prime}}{1+q^2}\mathcal{D}[b^\d_j]
A
\]
\be
+ \frac{q(\Gamma_2-\Gamma_1)}{1+q^2}\left(b^\d_j A b_{j^\prime} - \frac12\{b^\d_j
b_{j^\prime},A\}\right)\biggr\},
 \en{EQ11}
\be
H^{(0,0)}=\sum\limits_{j=1,2}\frac{\kappa_j}{1+q^2}(qb_j^\d+(-1)^jb_{j^\prime}^\d)^2(qb_j+(-1)^jb_{j^\prime})^2,
\en{EQh00}
where $j^\prime = 2$ for $j=1$ and vice versa.

We assume  that the parameters are chosen in such way that  mode $b_1$ has the
shortest evolution time and can be adiabatically eliminated. We consider two
particular limiting cases. In the first one both modes $a_{1,2}$ are carried by the
nonlinear media with the same nonlinearity, thus $\kappa_1=\kappa_2$. In the second
case  only   one of the two modes is nonlinear. As we shall see, there are
significant differences between these cases.

\begin{figure}[htb]
\begin{center}
\epsfig{file=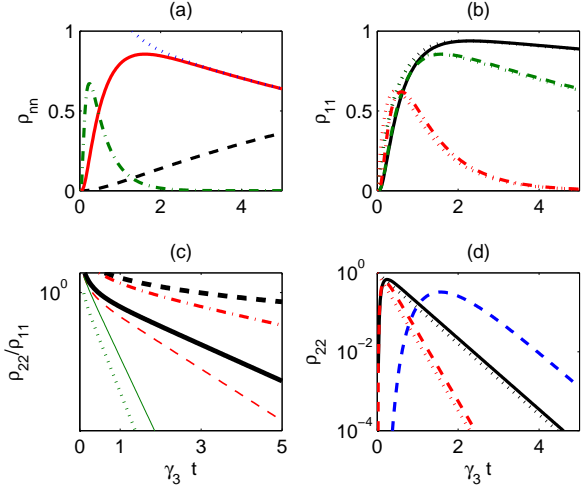,width=0.5\textwidth} \caption{(Color
online) (a) Dynamics of $\rho_{00}$ (dashed line), $\rho_{11}$
(solid line), $\rho_{22}$ (dash-dotted line) and average photon
number (dotted line)as given by the solution of Eq.~(\ref{MEQ});
$\Gamma=0.1\gamma_3$. (b) Dynamics of $\rho_{11}$ as given by
Eq.~(\ref{MEQ}) for the linear loss rate $\Gamma=0.01\gamma_3$,
$0.1\gamma_3$, $\gamma_3$ (solid, dashed and dash-dotted lines).
Dotted lines depict the solutions given by Eq.~(\ref{SPA}) for the
corresponding  values of $\Gamma$. (c) Dynamics of the ratio
$\rho_{22}/\rho_{11}$ for $\Gamma=0.1\gamma_3$, $\gamma_3$,
$10\gamma_3$ (thick solid, thin dashed, thin solid lines). Thick
dashed, dash-dotted and dotted lines correspond to the same values
of the linear rate, but with $\gamma_3$ being a hundred times
smaller. (d) Dynamics of $\rho_{22}$ as given by Eq.~(\ref{MEQ})
for the linear loss rate $\Gamma=0.1\gamma_3$ and
$\Gamma=\gamma_3$ (solid and dashed). Dotted lines depict the
solutions given by Eq.~(\ref{SPA}) for the corresponding values of
$\Gamma$. In all figures $\gamma_2=0.01\gamma_3$, and the solution
to Eq.~(\ref{MEQ}) is found for the initial coherent state with
the amplitude $z = 7$. }\label{fig3}
\end{center}
\end{figure}

\subsection{Two nonlinear side modes: generation of the single-photon state}
\label{sec3a}

Let us consider first the case of two nonlinear side modes $a_{1,2}$, when one has
$\kappa_1=\kappa_2\equiv\kappa$ and $\Gamma_1=\Gamma_2\equiv\Gamma$. This case was
considered before \cite{MS} as the scheme for the single-photon state generation.
Here we turn our attention to the important  problem of interplay between the
linear and nonlinear losses arising on the long evolution time scale. Also, we
outline some general features of dynamics produced by  the artificial nonlinear
losses.

In Ref. \cite{MS} it was shown that after the modes $a_0$ and
$b_1$ are emptied, the leading-order master equation for an
observable $A=A(t,b^\d_2,b_2)$ becomes
\be \frac{d A}{dt} \approx
i[H^{(0,0)},A] +
\Gamma\mathcal{D}[b^\d_2]A+\gamma_2\mathcal{D}[(b^\d_2)^2]A +
\gamma_3\mathcal{D}[(b^\d_2)^2b_2]A ,
\en{MEQ}
with the  decay rates
\be \gamma_2=\frac{4\kappa^2q^4}{(1+q^2)^4(\frac{4G^2}{\Gamma_0}+\Gamma)}, \quad
\gamma_3=\frac{4\kappa^2q^2(q^2-1)^2}{(1+q^2)^4(\frac{4G^2}{\Gamma_0}+\Gamma)}
\en{nlrates}
and the Hamiltonian
\be
H^{(0,0)}=\frac{\kappa(q^4+1)}{2(1+q^2)^2}n^2_2 + \left(\frac{\kappa
q^2}{(1+q^2)^2}-\frac{\kappa}{2}\right)n_2.
\en{H2}
We note that the symmetric coupling, i.e. $q=1$ leads to $\gamma_3=0$ (see Eq.
(\ref{nlrates})) and was considered recently in Ref. \cite{SM} in the context of
cold bosons in the triple-well trap. One of the notable results is that the
two-photon dissipation (with the rate $\gamma_2$) cannot be captured by the
mean-field approach, usually applied to systems of large number of bosons, though
the induced linear dissipation  is described by the mean-field model. Thus, quite
naturally, one can expect of such a  dissipation to generate  nonclassical states.

When the linear  and two-photon losses are absent ($\Gamma=0$ and $\gamma_2=0$),
the scheme described by Eq. (\ref{MEQ}) is able to generate the single-photon state
from a large amplitude  coherent input \cite{NlAbs,MS}. Indeed, the purity of the
output state can be easily estimated. The three-photon dissipation (with the rate
$\gamma_3$) leaves the following operators invariant: $A_{00} =
|0\rangle\langle0|$, $A_{01} = |0\rangle\langle1|$ and $A_{10}=A_{01}^\dag$. Since
for large times the density matrix $\rho$ involves only the Fock states $|0\rangle$
and $|1\rangle$, the following relations are valid: \be
\rho_{00}(\infty)=\rho_{00}(0), \; \rho_{10}(\infty)=\rho_{10}(0),\;
\rho_{11}(\infty)=\sum_{k=1}^\infty\rho_{kk}(0).
\en{EQ15} Eq. (\ref{EQ15})  gives for the coherent input state $\rho(0) =
|z\rangle\langle z|$ the following estimates: $\rho_{00}\approx e^{-|z|^2}$,
$\rho_{10}\approx z e^{-|z|^2}$ and $\rho_{11}\approx 1 - e^{-|z|^2}$. Thus, for
$|z|\gg1$ and $t\gg 1/\gamma_3$ the density matrix $\rho$ is exponentially close to
the single photon state $|1\rangle\langle1|$, i.e. $1-\mathrm{Tr}\{\rho^2\}\approx
2e^{-|z|^2} $.

Since neither of $g_j$ can be zero, the two-photon losses are also inevitable.
However, for a strong asymmetry, e.g. $|g_1|\gg |g_2|$, we get
$\gamma_3/\gamma_2=(q-1/q)^2\gg1$. Therefore the three-photon decay channel
dominates, resulting in  the single-photon state generation  from any coherent
input with a large amplitude \cite{MS}. For example, for $|q|=10$ the fidelity of
the single-photon state generation can exceed $99\%$. For such an asymmetric
coupling, in the absence of the linear losses, the resulting state is approximated
by the following one
\eqb |\textrm{out}\rangle\approx
\frac{1}{\sqrt{1+q^2}}\left[q|0\rangle_1|1\rangle_2-|1\rangle_1|0\rangle_2 \right],
\eqe
i.e. the photon is carried mostly by the weakly coupled mode $a_2$. The  linear
losses, however,    gradually degrade the generated state. For example, for the
linear loss rate an order of magnitude smaller than $\gamma_3$, the fidelity of the
single-photon state generation drops to $86\%$ \cite{MS}. The effect  of the linear
loss is smaller for smaller  times. However,  in that case a significant
multi-photon contribution to the state is still present  (Fig. \ref{fig3}(a)). The
multi-photon component can be a significant obstacle  for using the generated state
in the quantum cryptography protocols, where a presence of the two-photon component
in the signal can be used for security breaking \cite{LO}.

For small linear losses one can obtain a simple estimate of the parameters range
when   there is a significant probability of the single-photon state simultaneously
with a small probability of  the multi-photon components in the signal. We note
that in the Fock state  basis only the elements $\rho_{n,n+k}$ with a fixed $k$ of
the density matrix are coupled by Eq. (\ref{MEQ}). The diagonal elements satisfy
the following equation \be \frac{d\rho_{nn}}{dt} = \sigma_n\rho_{n+2,n+2}
+\beta_n\rho_{n+1,n+1} -\Gamma_n\rho_{nn}, \en{EQD} where $\sigma_n =
(n+2)(n+1)\gamma_2,\quad \beta_n = (n+1)(\Gamma +n^2\gamma_3)$ and
$\Gamma_n=n[\Gamma +(n-1)\gamma_2+(n-1)^2\gamma_3]$. Assuming that the mode $b_2$
is initially in the two-photon Fock state, from Eq. (\ref{EQD}) one easily obtains
\begin{eqnarray}
&&\rho_{22}=\exp{\left\{-2(\Gamma+\gamma_2+\gamma_3)t\right\}}, \nonumber\\
&&\rho_{11}=\frac{\gamma_3+\Gamma}{\gamma_2+\gamma_3+\Gamma/2} \nonumber\\
&&  \times\left(\exp\left\{-\Gamma t \right\}-
\exp\left\{-2(\Gamma+\gamma_2+\gamma_3)t\right\}\right) \label{SPA}
\end{eqnarray}
and $\rho_{00}=1-\rho_{11}-\rho_{22}$.

It is interesting that the simple estimate (\ref{SPA}) gives a fairly accurate
approximation of the single-photon generation fidelity (i.e. $\rho_{11}$) for times
when the initially coherent state of  mode $b_2$ approaches the single-photon level
(Fig. \ref{fig3}(b)). The approximation is quite precise even for the linear decay
rate approaching  $\gamma_3$. One can see that the position of the fidelity maximum
is also captured quite precisely. Apart from the initial time interval when the
state is far from the single-photon state, the estimate (\ref{SPA}) describes quite
accurately the behavior of the two-photon component, $\rho_{22}$ (Fig.
\ref{fig3}(d)).

Eq. (\ref{SPA}) allows to estimate  $\rho_{22}/\rho_{11}$ at the time when the
output state attains its maximal fidelity of the single-photon component. This
latter time, $T$, is defined by the condition $d\rho_{11}/dt=0$ and reads \eqb
T=\frac{1}{2\gamma_2+2\gamma_3+\Gamma}\ln\left(1+\frac{\gamma_2+\gamma_3}{\Gamma}\right).
\eqe This estimate of the fidelity maximum gives a very simple
result for the ratio \eqb
\frac{\rho_{22}(T)}{\rho_{11}(T)}=\frac{\Gamma}{\Gamma+\gamma_3}.
\eqe
This result seems rather disappointing, since for having a small ratio
$\rho_{22}/\rho_{11}$ at the maximal single-photon fidelity  one needs to have the
linear loss rate much smaller then the nonlinear one, $\Gamma\ll\gamma_3$. However,
if we relax the condition of having the maximal possible single-photon fidelity,
then even a strong  linear loss does not eliminate the applicability of our scheme
for the single-photon state generation. Eq. (\ref{SPA}) shows that $\rho_{22}$
decays faster than in the linear case, and $\rho_{11}$ decays much slower. Even for
the linear loss rates  exceeding $\gamma_3$, our scheme is an advantage if compared
to the simple linear attenuating of a coherent state (Fig. \ref{fig3}(c)).

\subsection{The nonclassical states generation with one nonlinear and one linear side modes}
\label{sec3b}

Now let us consider the asymmetrical case, when only one of the side modes is
carried by nonlinear (and, consequently, lossy) media. We assume that $\kappa_2=0$,
$\kappa_1\equiv\kappa$, and the rate of the linear loss is much less than that of
the nonlinear one, $\Gamma_2\ll\Gamma_1$. The nonlinear term of the Hamiltonian
(\ref{EQ8}), written in the $b$-basis, defines the nonlinear decay channels.
Expanding in terms of $b^\d_1$ and $b_1$ we get the operator coefficients (see Eq.
(\ref{EQ2})):
\be H^{(0,1)} = -\frac{2q\kappa}{(1+q^2)^2}(b^\d_2)^2 b_2, \quad H^{(0,2)} =
\frac{q^2\kappa}{(1+q^2)^2}(b^\d_2)^2. \en{EQ12} The coupling asymmetry defines the
resulting quantum state and the output channel (in this case,  the local mode $a_1$
or $a_2$).

\noindent \textit{The single-photon  state generation.--}
First, we consider the case of $q\ll 1$, which leads to the localization of the
coherent modes as follows $b_1 \approx a_2 +q a_1$ and $b_2 \approx -a_1 + qa_2$.
Note that, since  mode $b_1$ is empty, the r.h.s. of Eq. (\ref{EQ11}) simplifies in
this case to the diagonal form with the loss rate $\Gamma_{b_2} \approx
\Gamma_{1}+q^2\Gamma_2$ (note that the contribution from the linear losses are
negligible since $q^2\Gamma_2\ll\Gamma_1$). Resuming, when the modes $a_0$ and
$b_1$ are emptied, the leading-order master equation for an observable
$A=A(t,b^\d_2,b_2)$ becomes \be \frac{d A}{dt} \approx i[H^{(0,0)}\!\!,A] +
\Gamma_{b_2}\mathcal{D}[b^\d_2]A+\gamma_2\mathcal{D}[(b^\d_2)^2]A +
\gamma_3\mathcal{D}[(b^\d_2)^2b_2]A , \en{EQ13} with $H^{(0,0)}= \kappa (b^\d_2)^2
b_2^2$ and the loss rates
\[\gamma_2 \approx
\frac{16q^4 \kappa^2}{4G^2/\Gamma_0 + \Gamma_1}, \quad \gamma_3   
\approx \frac{16q^2\kappa^2}{4G^2/\Gamma_0 + \Gamma_1},\] i.e. $\gamma_2/\gamma_3
\approx q^2\ll1$, which is the condition for the single-photon state generation
\cite{NlAbs}. This scheme is similar to our previous proposal \cite{MS}, considered
in section \ref{sec3a}, but involves just one nonlinear side mode. The
single-photon state is still localized in the nonlinear mode ($a_1$). Below we also
present the schemes which generate quantum states in the \textit{linear} output
modes of the system (whether such a scheme exists for the single-photon generation
is yet unknown).

The  evolution interval  for the generation of single-photons  is quite similar as
in Fig. \ref{fig3} and  given by the condition $1/\gamma_3\ll t\ll 1/\Gamma_1$. The
sufficient conditions (used in the two reductions and the efficiency condition) for
the single-photon generation  read:
\be
\frac{G}{\Gamma_0}\ll1; \; \kappa,\,q\Gamma_1\ll \frac{G^2}{\Gamma_0}; \;
\frac{G^2\Gamma_1}{\Gamma_0}\ll (q\kappa)^2,
\en{EQ16}
where the last condition   is needed for  efficiency of the generator. The
consistency of conditions (\ref{EQ16}) requires that ${\Gamma_1\Gamma_0}/{G^2}\ll
q^2$, i.e. the efficiency of the generator is limited by the   losses in  the
nonlinear Kerr mode.
\medskip

\noindent \textit{The phase  state generation.--} Let us now
consider the case of $q\gg1$, which also leads to the nonclassical states
generation. In this case the scheme emulates the two-photon absorption, as is seen
from Eq. (\ref{EQ12}). Therefore, for an initial coherent state, $|z\rangle$ with
$|z|\sim 1$, the output density matrix of mode $b_2$ is  very close to the phase
state $|0\rangle + e^{i\arg(z)}|1\rangle$ \cite{TFAb,TFComm,TFRepl}. Note that the
output mode is still given by $b_2$, however now it is localized in the
\textit{linear mode}, $b_2 \approx a_2 - q^{-1} a_1$. We have the same form of the
approximate master equation (\ref{EQ13}) but now with the reduced Hamiltonian
$H^{(0,0)} \approx (\kappa/q^4)(b^\d_2)^2b^2_2$,  $\gamma_3 \approx
\frac{4\kappa^2\Gamma_0}{q^6G^2}$ and   $\gamma_2 \approx
\frac{\kappa^2\Gamma_0}{q^4G^2}$,    with $\gamma_3/\gamma_2\approx q^{-2}\ll1$,
where we have assumed that the linear mode $a_2$ has  negligible losses. Moreover,
since the operating mode $b_2$ is almost localized in the linear mode $a_2$ the
linear loss rate of this mode is almost equal to that of  mode $a_2$, i.e. we have
from Eq. (\ref{EQ11}) in this case: $\Gamma_{b_2} \approx \Gamma_2 + \Gamma_1/q^2$
(the cross-term vanishes since the mode $b_1$ is empty). The necessary conditions
on the scheme parameters in this case read
\be \frac{G}{\Gamma_0}\ll1,\;
qG\sqrt{\frac{\Gamma_1}{\Gamma_0}}\ll \kappa \ll \frac{G^2}{\Gamma_0}, \en{EQ17}
with a similar consistency condition as for the single-photon generator, i.e.
${\Gamma_1\Gamma_0}/{G^2}\ll 1/q^2$.

\subsection{Generation of the multi-photon nonclassical states}
\label{sec3c}

Usefulness of the three-mode scheme for generating of the nonclassical states is
not limited to generation of  the single-photon and phase states. As it was
mentioned in the Introduction, the nonlinear decay can be designed to produce an
arbitrary quantum state. In particular, the nonlinear dissipation with the Lindblad
generator of the type as in Eq. (\ref{MEQ}), i.e. $f(n_2)b^\d_2$, where $f(x)$ is a
positive definite function, in the absence of the linear losses can lead to the
stationary superposition of any given number of Fock states, or even to the
``combing" of the initial coherent state by retaining only the components with
either even or odd photon numbers \cite{mich}.

Moreover, for a sufficiently small evolution time even Eq.
(\ref{MEQ}) can lead to strongly nonclassical states. In
generation of such states   our scheme can be much more robust
with respect to the linear losses than in generation of the
single-photon state. It is easy to see that the modal dynamics in
the presence of nonlinear loss can be fairly impervious to the
linear decay for times much smaller than the characteristic time
of both  linear and nonlinear decay. This can be shown by invoking
an argument from the quantum jumps theory. Let us assume, for
simplicity, that the evolution is purely dissipative (the
Hamiltonian  is zero). Then the quantum Monte-Carlo  trajectory,
$|\psi(t)\rangle$, undergoes random jumps described by the
Lindblad operator $L$  \cite{ME}:
\[|\psi(t)\rangle \rightarrow L|\psi(t)\rangle.\]
The probability of the jump is proportional
$\langle\psi(t)|L^{\dagger}L|\psi(t)\rangle$. Obviously, the state components with
the high  photon numbers will be ``jumping down" via the nonlinear decay described
by Eq. (\ref{MEQ}) with much larger rate than that of the linear decay. For
instance, it was shown   that for $L=b_2^2$  quite different initial states decay
to the stationary state practically in the  same time  \cite{DodMiz}.

\begin{figure}[htb]
\begin{center}
\epsfig{file=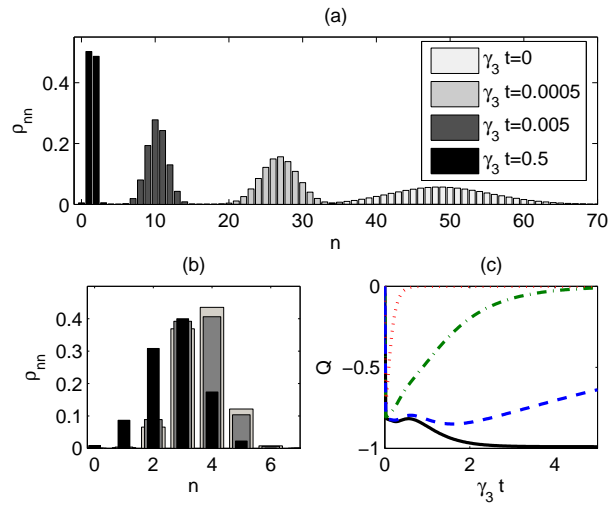,width=0.5\textwidth} \caption{(Color online) (a) Evolution
of the photon-number distribution for small losses $\Gamma=0.1\gamma_3$. (b) The
photon-number distribution at $\gamma_3 t=0.3$; light gray, dark gray and black
bars correspond to $\Gamma=0$,$0.1\gamma_3$ and $\gamma_3$. (c) The corresponding
Mandel parameter $Q$, given by Eq. (\ref{mandel}), for $\Gamma=0$, $0.1\gamma_3$,
$\gamma_3$ and  $10\Gamma_3$ (solid, dashed, dash-dotted and dotted curves). We use
the solution of Eq. (\ref{MEQ}) for the initial coherent state with the amplitude
$z=7$. Here $\gamma_2=0.01\gamma_3$. }\label{fig4}
\end{center}
\end{figure}

Thus, one can expect that a classical state can be transformed into a nonclassical
one by the scheme described by   Eq. (\ref{MEQ}) in a short evolution time  for
which the linear losses are still negligible. Indeed, for two nonlinear side modes,
when $\gamma_3\gg\gamma_2$, one should expect a significant change of the quantum
state to occur for the evolution time on the order
\eqb t\sim \left(\gamma_{3}\langle n_2^2\rangle \right)^{-1}.
\label{time estimate}
\eqe
Fig. \ref{fig4}(a) confirms this prediction  (there the expected significant change
occurs  at $\gamma_3 t\approx 0.0004$, i.e. exactly at the estimate (\ref{time
estimate})). For such times the average number of photons drops by nearly two
times. Naturally, the linear decay rate should be of several orders of magnitude
larger than the nonlinear one   to play any significant role in this case.

We also note that even for larger times  and a smaller average photon number the
generated state is quite robust with respect to the linear losses. For example, for
the states with three or four photons on average, the linear losses with the rate
even comparable with $\gamma_3$  do not destroy the nonclassicality (Fig.
\ref{fig4}(b,c)). It is interesting that for moderate linear losses the few-photon
states, generated this way, can be very interesting.  One can obtain, for instance,
a nearly equal mixture of the one and two-photon components, Fig. \ref{fig4}(a).

The nonclassicality of the generated state can be  illustrated by using  the Mandel
parameter,
\eqb
Q=\frac{\langle n_2^2\rangle - \langle n_2\rangle^2}{\langle n_2\rangle} -1,
\label{mandel}
\eqe
which is zero for a coherent state.  It is seen from Fig. \ref{fig4}(c) that for
the initial coherent state the generated state is manifestly a sub-Poissonian one
during all $t > 0$. In the absence of  linear losses one can distinguish three
stages of the $Q$-parameter evolution. First, it sharply drops (for the time-scale
given by the estimate (\ref{time estimate})). Then, when the state has a sharp
few-photon maximum, the Mandel parameter oscillates. Finally, it   slowly
approaches $Q_F=-1$,  i.e. its value for a Fock state.

For the small and moderate linear loss rate, $\Gamma \leq \gamma_3$, only the third
stage   appears to be affected: the Mandel parameter rises slowly towards zero. As
long as the   linear decay rate remains larger then  the time-scale given by the
estimate (\ref{time estimate}), the first stage is practically unaffected. During
this stage, the quantum state undergoes strong photon-number squeezing.

Thus the considered scheme can serve as an efficient and robust practical source of
the  sub-Poissonian nonclassical states. Obviously, all kinds of set-ups mentioned
above   serve for this purpose. For example,  a few centimeters of the three-core
fiber with the realistic   nonlinearity and   linear losses  is enough for reaching
$Q\sim 0.1$ \cite{MS}.

We conclude that the multi-core fiber realizations of the scheme described in
section \ref{sec2} can be quite useful for generating of the multi-photon
nonclassical states despite the fact that a strong Kerr nonlinearity is usually
accompanied by a strong linear absorption.  Linear losses can be dealt with more
efficiently in the scheme with the individual damped emitters (quantum dots, atoms,
color defects etc.) acting as a common loss reservoir for the light modes.
Moreover, the EIT-like schemes exhibiting a very strong Kerr-nonlinearity
\cite{kerr} might be a promising candidate for realization  of our setup.


\section{The entangled state generation by the four-mode coupler}
\label{sec4}

Let us now explore  a four-mode  generalization of our scheme.  We assume that a
strongly lossy mode $a_0$ is coupled to three other modes, one from which is a
nonlinear mode, $a_3$, and two other are linear ones, $a_{1,2}$. We consider the
Hamiltonian
\be
H_4 = G\left(a^\dag_0\sum_{\ell=1}^3 g_\ell a_\ell  + a_0\sum_{\ell=1}^3 g_\ell
a^\dag_\ell\right) + \kappa (a^\dag_3)^2a_3^2,
\en{EQ18}
where   the coupling coefficients $g_\ell$ are real and  normalized $\sum_\ell
g^2_\ell=1$. Moreover, for  specific localization of the coherent modes (see below)
the $g_\ell$ are assumed to be strongly asymmetric: $g_1=g_2 = \epsilon\ll1$ and
$g_3 = 1-\mathcal{O}(\epsilon^2)$. The master equation for the observable $A$ of
the system reads
\be
\frac{d A}{dt}= i[H_4,A] +
\Gamma_0\mathcal{D}[a^\dag_0]A+\Gamma_3\mathcal{D}[a^\dag_3]A,
\en{ME}
where we neglect the  losses in the linear modes $a_{1,2}$ and assume that the
strong dissipation of mode $a_0$ has the shortest time scale, i.e. $\Gamma_{0} \gg
G, \Gamma_3, \kappa$. To perform the steps which lead us to the second reduction,
Eq. (\ref{EQ7}), we need to introduce  a coherent basis $b_\ell$ given by the
unitary matrix  $U$ with the constraint  $U_{\ell,3}=(g_1,g_2,g_3)$. This matrix
can be chosen as follows
\be
\left(\begin{array}{c} b_1\\ b_2\\b_3 \end{array}\right) = \left(\begin{array}{ccc}
\sqrt{1-g_1^2} & -\frac{g_1g_2}{\sqrt{1-g^2_1}} & -\frac{g_1g_3}{\sqrt{1-g^2_1}}\\
0& \frac{g_3}{\sqrt{1-g^2_1}} & -\frac{g_2}{\sqrt{1-g^2_1}}\\
g_1 & g_2 & g_3\end{array}\right)\left(\begin{array}{c} a_1\\
a_2 \\ a_3 \end{array}\right).
\en{EQ19}
In the strongly asymmetric case,  the rotation (\ref{EQ19}) can be approximated as
$b_\ell = a_\ell - \epsilon a_3+\mathcal{O}(\epsilon^2)$, $\ell=1,2$ and $b_3 = a_3
+ \epsilon(a_1 +a_2) + \mathcal{O}(\epsilon^2)$. In this case, the adiabatic
procedure described in Section II eliminates the lossy mode $a_0$ 
(the first stage) and then the coherent mode $b_3$ (the second stage). For the
evolution times $t\gg 1/(\Gamma_3 + 4G^2/\Gamma_0)$, i.e. when mode $b_3$ is
already in the vacuum state, the  two coherent modes $b_{1,2}$ possess two channels
of decay, as given by Eq. (\ref{EQ7}). These channels correspond to the two
Lindblad generators $H^{(0,1)}$ and $H^{(0,2)}$   obtained by the expansion (as in
Eq. (\ref{EQ2})) of Hamiltonian (\ref{EQ18}) in powers of $b^\dag_3$ and $b_3$. In
the leading order approximation we have:
\[
H^{(0,1)} = -\epsilon^3\kappa(b^\dag_1+b^\dag_2)^2(b_1 +b_2) +
\mathcal{O}(\epsilon^5),\] \be H^{(0,2)} =
\epsilon^2\kappa\left[(b^\dag_1)^2+(b^\dag_2)^2 +b^\dag_1
b^\dag_2\right]+\mathcal{O}(\epsilon^4),
\en{EQ20}
hence the corresponding decay rates in Eq. (\ref{EQ7}) (with $\Gamma = \Gamma_3 +
4G^2/\Gamma_0$) read $\gamma_2 \approx 4\epsilon^4\kappa^2/(\Gamma_3 +
4G^2/\Gamma_0)$ and $\gamma_3 \approx 4\epsilon^6\kappa^2/(\Gamma_3 +
4G^2/\Gamma_0)$. The reduced two-mode Hamiltonian is $H^{(0,0)} =
\epsilon^4\kappa(b^\dag_1 + b^\dag_2)^2 (b_1+b_2)^2$.

The output states of this four-mode scheme are  localized in the linear modes
$a_{1,2}$. Note that  the coupling to the lossy nonlinear mode $a_3$ induces the
coherent linear losses with the Lindblad generator $b_1^\dag +b_2^\dag$  and a
reduced loss rate $\Gamma_L \equiv \epsilon^2\Gamma_3$, which fact can be easily
established by expansion of the term $\Gamma_3\mathcal{D}[ a_3^\dag]=
\Gamma_3\mathcal{D}[\sum_k U_{k3}b_3^\dag]$ similar as in Eq. (\ref{EQ11}).  For
below, we can neglect  the decay channel $H^{(0,1)}$, with the decay rate
$\gamma_3$, as compared to the channel $H^{(0,2)}$ and the linear losses.

The sufficient conditions for the adiabatic elimination of the lossy modes and the
condition of the scheme efficiency read
\be
\Gamma_0\gg G, \Gamma_3, \kappa;\quad \epsilon^2 \kappa \ll \Gamma_3 +
\frac{4G^2}{\Gamma_0} \ll \epsilon \kappa,
\en{EQ27}
with the simple unrestrictive consistency condition  $\epsilon \ll 1$ (cf. with
that for the three-mode scheme of section \ref{sec3}). In Eq. (\ref{EQ27}) the
rightmost inequality  is the requirement of the scheme efficiency (the artificial
nonlinear loss rate $\gamma_2$ is stronger than the induced linear loss in
$a_{1,2}$), while the middle one together with  the left one (already mentioned
above) are sufficient conditions for the adiabatic elimination of the lossy modes.
Under these conditions, the strongest nonlinear dissipation, with $H^{(0,2)}$, the
linear losses term $\Gamma_L\mathcal{D}[b_1^\dag +b_2^\dag]$ and the reduced
Hamiltonian $H^{(0,0)}$ together govern   the evolution of observables.

If the liner losses were absent, the nonlinear losses  term would dominate the
Hamiltonian part, since $\gamma_2\gg \epsilon^4 \kappa$, i.e. $\kappa\gg \Gamma_3 +
4G^2/\Gamma_0$, which is satisfied due to conditions (\ref{EQ27}). To find out what
structure the output density matrix has, let us simplify the Lindblad generators by
performing a rotation to the coherent basis:
\be
 c_{\pm} = \frac{b_1 \pm b_2}{\sqrt{2}}
=\frac{a_1\pm a_2}{\sqrt{2}} - \epsilon\sqrt{2} a_3 +\mathcal{O}(\epsilon^2).
\en{EQ21}
We have  $H^{(2,0)} = \epsilon^2\kappa(3c_+^2 - c_-^2)/2$ and  $H^{(0,0)} =
4\epsilon^4\kappa (c^\dag_+)^2c^2_+$. Then, up to the order $\epsilon^4$ the master
equation (now written for the density matrix, in the coherent basis)  reads
\be
\frac{d \rho}{dt}= -i[H^{(0,0)},\rho] + \gamma_2\mathcal{D}[(3c_+^2 - c_-^2)/2]\rho
+ 2\Gamma_L\mathcal{D}[c_+]\rho.
\en{EQ24}
The nonlinear loss term has the null subspace given by the following expansion in
the Fock basis of the coherent modes
\be
|\textrm{null}\rangle = \sum_{n,m}f(n)g(m)|n,m\rangle,
\en{Null}
where each $n$ and $m$ take only even or odd values (independently), while  $f(k) =
\frac{f_0}{3^{k/2}\sqrt{k!}}$   and $g(l) = \frac{g_0}{\sqrt{l!}}$. Note that the
coherent state $|\alpha,\sqrt{3}\alpha\rangle$  belongs to this null subspace.
However, only the following Fock basis vectors
\be
|0,0\rangle,\; |0,1\rangle,\; |1,0\rangle,\;  |1,1\rangle,
\en{DS}
belong to the joint null subspace of $H^{(2,0)}$ and $H^{(0,0)}$.

We will further concentrate on the realistic case with the conditions (\ref{EQ27})
satisfied, i.e. when there is also the linear loss  term and the decay rates are
ordered as follows
\[
\gamma_3\ll \epsilon^4\kappa \ll  \Gamma_L \ll \gamma_2.
\]
In this case the output density matrix can be found only numerically.  Since the
initial dynamics (i.e. effecting the two adiabatic reductions) almost leaves  the
non-dissipated modes unaffected, the state of these modes is close to the initial
state. Hence,  one can pass to simulating the resulting approximate master equation
(\ref{EQ24}) instead of that for the four coupled modes. While   the average values
can be found numerically without much effort for quite large amplitudes of the
initial coherent state, the entanglement measure (see below) requires knowledge of
the whole underlying density matrix. This limits the efficiency of the numerical
simulations and  we are able to simulate thoroughly   only the small amplitude case
of $|\alpha_j|\sim 1$.

\begin{figure}[ht]
\begin{center}
\epsfig{file=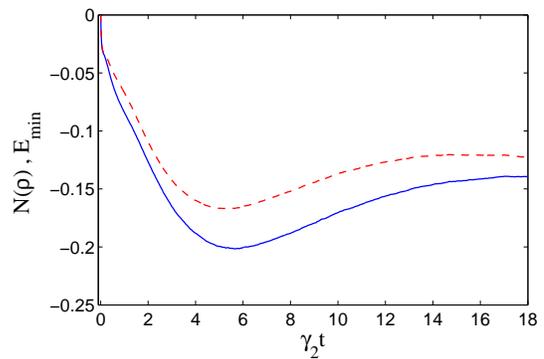,width=0.4\textwidth} \caption{(Color online) The entanglement
witness (solid), the sum of the negative eigenvalues of the partial transpose of
the density matrix of modes $a_{1,2}$ restricted to  Fock indices
$n\in\{0,...,5\}$ in each mode, and the corresponding lowest negative eigenvalue
(dashed). Here the parameters read $\epsilon = 0.1$, $\kappa = 30$, $\Gamma_0 =10$
and $\Gamma_3 = 2$ which give $\gamma_2 = 0.15$, $\epsilon^4\kappa = 0.012$ and
$\Gamma_L = 0.024$. We have used the following coherent state
$|-i\alpha,0.9\alpha\rangle$ as the input state with $\alpha=1.6$ (using another
input state with the similar average number of photons does not change the
picture). The figure is an average over 17000 quantum trajectories. }
\label{FG3}
\end{center}
\end{figure}

Fortunately, considering  the small amplitude case     is sufficient for concluding
on  the general case. This is because the nonlinear Lindblad term of the master
equation (\ref{EQ24}) admits some integral invariants, which are also invariants of
the Hamiltonian dynamics. Indeed, since in the absence of the linear losses the
photons are removed by pairs, the Hilbert space of   each of the coherent modes is
effectively divided into two subspaces, involving only either even or odd Fock
number states. Hence, introducing the identity operators in each parity subspace,
$I_{ev}$ and $I_{odd}$, the above mentioned integral invariants are given by taking
the trace of four possible products $I_{\sigma_1}\otimes I_{\sigma_2}$,
$\sigma_{1,2}\in\{ev; odd\}$, with the density matrix. For $t\to\infty$ they define
the following diagonal output matrix elements $\rho_{00,00}$, $\rho_{10,10}$,
$\rho_{01,01}$, $\rho_{11,11}$. For the coherent pure input state
$|\alpha_1,\alpha_2\rangle$ (in the modes $c_{\pm}$) we get \be
\rho_{nm,nm}(\infty) = e^{-|\alpha_1|^2-|\alpha_2|^2}
F^{(n)}(|\alpha_1|^2)F^{(m)}(|\alpha_2|^2),
\en{MatEl} where
\[
F^{(0)}(|\alpha|^2) = \cosh(|\alpha|^2),\quad F^{(1)}(|\alpha|^2) =
\sinh(|\alpha|^2).
\]
We have verified numerically that for $|\alpha|^2\ge 2$ the two functions
$F^{(i)}(x)$ differ by less than 4\%, i.e.  all $\rho_{nm,nm}$ are very close to
each other for such arguments. Noticing that the output density matrix is close to
a pure state (and not proportional to the unit matrix) we conclude that  the output
state is an entangled state for any initial coherent state and small linear losses.

\begin{figure}[ht]
\begin{center}
\epsfig{file=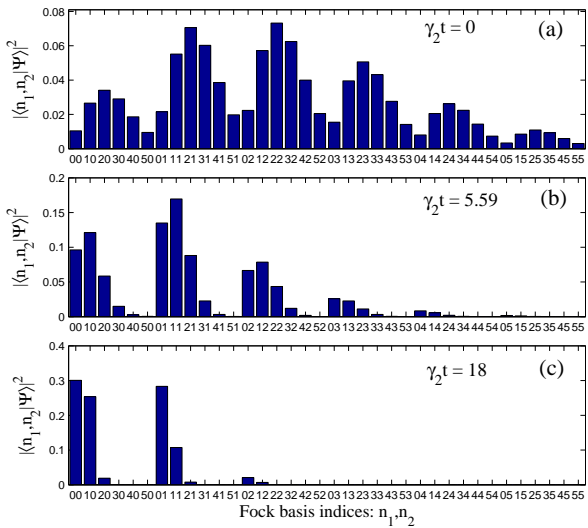,width=0.45\textwidth} \caption{(Color online) We give the
structure of the eigenstate of the density matrix corresponding to the maximal
eigenvalue  (i.e. the maximal probability) of the density matrix for three
different propagation times, $\gamma_2t=0$ (a), $\gamma_2t=5.95$ (b) and
$\gamma_2t=18$ (c). Here   the abscissa is the ordered   Fock basis indices
$n_1n_2$ of the state amplitude. The parameters are as in Fig. \ref{FG3}. }
\label{FG4}
\end{center}
\end{figure}

The predictions were checked with the   quantum Monte-Carlo simulations of the
master equation (\ref{EQ24}) in the local basis $a_{1,2}$ using  the ``quantum
jumps'' method \cite{ME}. The numerically obtained output  density matrix indeed
describes an entangled state of the two linear local modes $a_1$ and $a_2$, see
Fig. \ref{FG3}. We use the negativity  as a measure of the entanglement
\cite{Negat,Peres}, but  with the minus sign in front (for convenience of
comparison with the smallest negative eigenvalue). It is equal to  the sum of all
negative eigenvalues of the partially transposed with respect to the local modes
$a_{1,2}$ density matrix, $\rho^{PT}$: $\mathcal{N}(\rho)\equiv
\sum_{E<0}E(\rho^{PT})$ (due to technical reasons, in calculation of the negativity
the expansion over the Fock states  was truncated at  $n=5$,  however,   this turns
out to be sufficient to capture negative eigenvalues \cite{FN}, moreover, the
higher Fock number states are populated significantly only for the relatively short
time). The output density matrix turns out to be close to a pure state (for a pure
input state): the largest eigenvalue of the output density matrix drops during the
evolution only to $0.85$ at $\gamma_2t=18$ in Fig. \ref{FG3}. The significant
elements of the corresponding eigenstate of the density matrix involve the trivial
null states given by Eq. (\ref{DS}) and few Fock states with higher population of
the local modes $a_{1,2}$, see Fig. \ref{FG4}. Our simulations show that this seems
to be the only outcome for the output density matrix at least for  small amplitudes
of the coherent initial states. In view of the above arguments, this result must be
expected also for arbitrary amplitudes.

\section{Conclusion}

We have shown that it is possible to generate the nonclassical states of light by
emulation of the  nonlinear absorptions using the usual Kerr-nonlinearity, linear
mode coupling and the strong linear losses. We have concentrated on the potential
capabilities of the scheme to generate various quantum states in  the simplest
possible setups of three and four coupled modes. We have found that, additionally
to the single-photon states, also the phase states and the entangled states can be
generated by our method. It is interesting to observe, that though our scheme
depends significantly on the Kerr nonlinearity (which must be present at least in
one of the modes) the output states can be localized in the linear modes, as in the
case of the phase states and the two-mode entangled states. Though, we  have not
been able to present  a similar scheme for the single photon states, there is some
hope that such a scheme may also exist. Having   the output state localized  in the
linear modes of the system prevents its subsequent degradation   by the destructive
linear losses, accompanying any known strong Kerr nonlinearity, but quite small in
the linear media. We believe that our scheme can be realized in a practical setup
with the current technology. Some perspective candidates are the optical fiber
coupler or the multi-core fiber, the electromagnetically induced transparency
media, where  the ``giant" Kerr nonlinearity has been demonstrated \cite{kerr}.

\acknowledgements
 V.S.S. acknowledges the financial support by the CNPq and FAPESP (2010/16858-6)
of Brazil. D.M. acknowledges the financial support by the BRFFI of
Belarus.


\begin{thebibliography}{99}
\bibitem{FN} This is clear from the fact that, for a Hermitian
matrix, if any of the principal submatrices, i.e. consisting of the same subset of
rows and columns,  has negative eigenvalues, then the full matrix also has negative
eigenvalues.
\end{thebibliography}
\end{document}